\documentclass{article}
\usepackage[english]{babel}
\usepackage[T2A]{fontenc}
\usepackage[cp1251]{inputenc}
\usepackage[dvips]{graphicx}
\usepackage{amsmath}
\usepackage{amssymb}
\usepackage{latexsym}
\begin{document}
\rm
\let\sc=\sf
\begin{flushright}
Journal-Ref: Astronomy Letters, 2010, Vol. 36, No. 11, pp. 808–815
\end{flushright}
\begin{center}
\LARGE {\bf Modulation of Circumstellar Extinction in a Young Binary System
with a Low-Mass Companion in a Noncoplanar Orbit}\\

\vspace{1cm}
\Large {\bf V. P.\,Grinin$^{1,2}$, T. V.\,Demidova$^{1,2}$, N. Ya.\,Sotnikova$^1$}

\normalsize \vspace{5mm}
1 - Sobolev Astronomical Institute, St. Petersburg State University, Universitetskii pr. 28, St. Petersburg, 198504 Russia,

2 - Pulkovo Astronomical Observatory, Russian Academy of Sciences, Pulkovskoe shosse 65, St. Petersburg,
                                         196140 Russia, \\
\end{center}
Received April 10, 2010
\normalsize
\begin{abstract}
The cyclic activity model of a young star with the low-mass secondary component ($q$ = $M_2/M_1 \leq$
0.1) accreting a matter from circumbinary disk is considered. It
is assumed that the orbit is circular and the disk and orbital
planes are non-coplanar. Sets of hydrodynamics models of such a
system have been calculated by the SPH method and then the
variations of the circumstellar extinction and phase light curves
were determined. The calculations showed that depending on the
model parameters and orientation of the system in regards to an
observer the different in shape and amplitude light curves can be
observed. An important property of the considered models is also
the dependence of the mass accretion rate onto the components on
the phase of the orbital period. The results of the calculation
can be used for analysis of the cyclic activity of UX Ori stars
and young stars with the long-lasting eclipses.

Key words: \emph{young binary systems, extinction modulation, activity cycles of UX Ori stars}.
\end{abstract}
\clearpage

\clearpage
\large
\newpage
\section{INTRODUCTION}

 In the previous papers of this series (Sotnikova
and Grinin 2007; Demidova et al. 2010a, 2010b), we
studied the behavior of the circumstellar extinction
and brightness in young binary systems accreting
matter from the remnants of a protostellar cloud.
This problem was initiated by the observations of
large-scale activity cycles in UX Ori stars (Grinin
et al. 1998) and by the observations of anomalously
long eclipses in some young stars (Cohen et al. 2003;
Barsunova et al. 2005; Nordhagen et al. 2006; Grinin
et al. 2008), which cannot be explained in terms of
classical models of eclipsing binary systems. Sotnikova
and Grinin (2007) and Demidova et al. (2010a,
2010b) showed that the streams of matter and density
waves produced in young binary systems by the
orbital motion of their companions can cause (at low
inclinations of the orbital plane to the line of sight)
various (in shape and duration) periodic fadings of the
primary component. One of the main conditions in the
computations was coplanarity of the orbital plane and
the circumbinary (CB) disk.

In this paper, we consider the more general case
where the orbit of the companion and the CB disk
are \emph{noncoplanar}. The prototype of such a binary is
the star $\beta$ Pictoris surrounded by an extended circumstellar
disk (Smith and Terrile 1984) seen almost
edge-on. Detailed studies of the disk image (Burrows
et al. 1995) showed that its inner region was
inclined by several degrees relative to the periphery.
This peculiarity is considered as circumstantial evidence
for the existence of a planet in the central
part of the disk whose orbit is inclined with respect
to the outer part of the disk (Burrows et al. 1995;
Mouillet et al. 1997; Larwood and Papaloizou 1997).
The fact that the disk of $\beta$ Pictoris is the first and
so far the only example of a comparatively young
($\approx 10^7$ yr) circumstellar disks tudied very thoroughly
and that evidence for the existence of a companion in
a noncoplanar orbit was found in this disk suggests
that such situations are encountered in nature not so
rarely. This circumstance served for us as an incentive
to perform computations whose results are presented
below.
\section{MODEL PARAMETERS AND THE COMPUTATIONAL\\ METHOD}
We consider a model of a binary system with a
low-mass companion moving in a circular orbit of
radius $a$ inclined with respect to the CB-disk plane
by angle $\theta$. In our computations, we took $\theta$ = 10$^\circ$.
The other parameters of the problem are: the orbital phase-averaged
 accretion rate onto both components
of the binary ($\dot M_a$), the orbital inclination to the line
of sight ($i$), and the angle between the projection of
the line of sight onto the orbital plane and the line
of nodes ($\phi$). The mass of the central star $M_*$ was
taken to be 2$M_\odot$, a value typical of many UXOri stars
(Rostopchina 1999). The orbital period is $P = 5$ yr.

\subsection{The Computational Method}

As in Artymowicz and Lubow (1996), we computed
the hydrodynamic model in the isothermal approximation.
This approach is justified, because, as our
computations show, not the entire CB disk but only
its comparatively small region near the inner boundary
makes a major contribution to the circumstellar
extinction variations in our models.

The hydrodynamic models of the binary under
consideration were computed by the SPH (Smoothed
Particle Hydrodynamics) method described in detail
by Sotnikova (1996). Let us briefly recall the main
features of the computations. In our implementation
of the SPH method, the smoothing length was
assumed to be constant and was specified in fractions
of the orbital radius: $h = 0.1\,a$. At each point, this
provided at least $30$ neighboring points, over which
the hydrodynamic quantities were averaged. The
parameter $c$, which characterizes the viscosity and
is the dimensionless speed of sound expressed in
units of the companion’s Keplerian velocity, was
taken to be $0.05$ (a “warm” disk). (Note that for the
values of $M_*$ and $P$ adopted above and at a stellar
temperature of $10^4K$ typical of UX Ori stars, this
speed of sound corresponds to a gas temperature about
$100 K$.) The number of test particles used in our
SPH simulations was 6$\cdot$10$^4$. As a rule, the computations
were performed for several hundred periods.
For each model, we computed the column density
of test particles toward the primary component of
the binary as a function of time expressed in units
of the orbital period. The cross-sectional area of the
column $s$ was taken to be $2h \times 2h$. Our computations
showed this value of $s$ to be optimal for the solution of
our problem.

The technique for calculating the phase dependence
of the dust column density was described in
detail by Demidova et al. (2010a). Let us briefly recall
its main points. Our quantitative analysis of the computational
results begins with the removal of the trend
in the particle column density variations attributable
to the decrease in their number in the binary through
accretion onto its components. The initial binary relaxation
stage (with a duration of several tens of revolutions),
during which its matter distribution comes
to a stable state, is excluded from consideration. The
trend is modeled by a fifth-degree polynomial. Our
computations showed that this provided a satisfactory
removal of the trend for all of the models under
consideration. Subsequently, we passed from the test
particle column density $n(t)$ to the column density
of real dust grains $n_d (t)$ (for details, see Demidova
et al. 2010a). To reduce the influence of random
fluctuations when the phase dependences of $n$ are
computed, the current values of $n(t)$ are folded with
the orbital period for a time interval of $50$ binary
revolutions.

As in our previous papers, we assumed the primary
component to be the main source of optical radiation.
To determine the optical depth $\tau$ between the star and
the observer attributable to circumstellar dust, it is
necessary to find the column density of test particles
as a function of time and to determine the “mass” of
one such particle. For this purpose, we determined
the number of test particles accreting onto the binary
components at each instant of time (for more detail,
see Sotnikova and Grinin 2007). These values were
averaged over the orbital phases and their sum was
compared with the mean accretion rate onto both
components adopted in the model, $\dot M_a$. The “mass”
of a single test particle $m_d$ was determined from the
formula $m_{d}= P\,\dot M_a/N_a$, where $P$ is the orbital period
and $N_a$ is the total number of test particles accreting
onto both components in one binary revolution. In
our computations below, we took $\dot M_a$ to be $10^{-9}$ and
$10^{-10}\, M_{\odot}$. As our computations showed, these
values of the accretion rate are sufficient to produce a
high-amplitude brightness modulation in the primary
component.

Having determined the mass of a single test particle,
we obtain the matter column density in $g/cm^2$.
To calculate the optical depth of the dust on the line
of sight, we should specify the opacity $\kappa$ per gram of
matter. As in our previous papers, below we adopted
an average (for the interstellar medium) dust-to-gas
ratio of $1:100$ and $\kappa = 250$ $cm^2/g$ typical of circumstellar
extinction in Johnson’s $B$ photometric
band (Natta and Whitney 2000).

The intensity of the radiation from young stars is
known to consist of two parts: the intensity of the
direct stellar radiation $I_{\ast}$ attenuated by a factor of $e^{- \tau}$
and the intensity of the radiation scattered by circumstellar
dust $I_{sc}$:$I_{obs}=I_{\ast}e^{-\tau} + I_{sc}$. The contribution
of the scattered light to the total radiation from young
stars typically does not exceed a few percent. Therefore,
below, when analyzing the pattern of variability
in the primary component, we took the intensity of the
scattered radiation to be zero. The light variations of
the primary component are expressed in magnitudes:
$\Delta m = -2.5\cdot\log{I_{obs}}$ ($I_{\ast}$ is taken as unity). Hence it
follows that $\Delta m \sim \tau$ , and since $\tau$ is proportional to
$\dot M_a$, $\Delta m \sim \dot M_a$. This relation allows the light curves
computed for one value of $\dot M_a$ to be easily recalculated
for other values of this parameter.

There is another useful relation that allows the
light curves computed for a period of $5$ yr to be recalculated
to a different value of $P$. It can be derived from
the following considerations: according to the above
formula for the test particle mass, $m_d$ is proportional
to the orbital period $P$. On the other hand, when
calculating $\tau$, we divide the number of particles in the
column $n(t)$ by its cross-sectional area $s$. The latter is
proportional to $a^2 \sim P^{4/3}$. Taking this into account,
we obtain $\Delta m \sim \tau \sim P^{-1/3}$.

\section{RESULTS}
\subsection{Model Distributions of Particles in a Binary}

Figures 1 and 2 show the distributions of test
particles computed by the SPH method for two models
with different component mass ratios: q = 0.1
(model 1) and q = 0.01 (model 2). The coordinate
system is chosen in such a way that the XY plane
coincides with the CB disk plane, the coordinate origin
coincides with the binary’s center of mass, and the
Y axis coincides with the line of nodes. We see that in
model 1, the inner region of the CB disk is strongly
deformed and inclined with respect of its periphery
due to the gravitational perturbations produced by the
companion’s orbital motion. We see from the binary’s
sections in the YZ and XZ planes presented in Fig. 1
that the inner disk region is inclined not only with
respect to the Y axis but also in the YZ section
(i.e., with respect to the X axis). In other words, the
inclination of the inner CB disk region does not
coincide closely with the orbital inclination. This
pattern of deformation of the CB disks tems from the
fact that in the models with an inclined orbit, the angular
momentum vector for the companion’s orbital
motion has two components, one of which is orthogonal
to the disk plane and the other coincides with
it and is directed (in the adopted coordinate system)
along the X axis. The latter component produces the
CB disk inclination in the YZ section. This pattern
of deformation of the inner disk in the models with
an inclined orbit was first described in the paper by
Mouillet et al. (1997) cited above, where the particle
dynamics was computed in the ballistic approximation.
Our computations showed that this peculiarity is
also retained in the hydrodynamic models computed
by taking into account the gas pressure and viscosity.

The mass of the companion in model 2 is a factor of 10
lower than that in model 1. Therefore, the CB disk
deformation is much weaker. However, in contrast
to model 1, the region within the orbit is filled with
matter to a greater extent and we see from Fig. 2
that this matter has the shape of a disk, whose plane
coincides neither with the orbital plane nor with the
CB disk plane. This peculiarity in the distribution of
matter within the companion’s orbit is attributable
to the same effect as the CB disk deformation in
model 1 described above. Since the accretion disks of
the binary components are formed from this matter,
they will also be inclined with respect of the orbital
plane, in which the primary and secondary components
revolve around the binary’s center of mass.

\subsection{Accretional Activity of the Binary Components}

Figure 3 shows the phase dependences of the accretion
rate onto the binary components computed for
models 1 and 2. Just as in the models with coplanar
orbits (Artymowicz and Lubow 1996; Bate and Bonnell
1997; Demidova 2009), the low-mass component
is the main accretor (because its orbit is close to
the CB disk). However, in contrast to these models,
the accretion rate onto the components in models 1
and 2 depends on the orbital phase even for a circular
orbit. In this case, the pattern of modulation of the
accretion rate depends on the mass of the companion.
The accretion rate onto the low-mass component in
model 2 (q = 0.01) has two maxima. They correspond
to times that are close to the times of the companion’s
passage through the line of nodes (but that do not
coincide with them). The picture is completely different
in model 1 with a more massive companion: the
accretion rate has only one maximum instead of two.
In this case, the accretion rates onto the primary and
secondary components change in antiphase.

\subsection{Model Light Curves}

Figures 4–7 show the light curves of the binary’s
primary component computed by the method described
above. The computations were performed for
several orbital inclinations to the line of sight and
four angles specifying the position of the line of nodes
relative to the observer: $\phi = 0^\circ$, $90^\circ$, $180^\circ$, $270^\circ$. 
(In the frame of reference adopted here, the angle $\phi = 0$
corresponds to the case where the line of sight is orthogonal
to the line of nodes and the observer in Fig. 1
is located to the right; the angles are measured in the
direction of rotation of the binary). As in our previous
papers, the range of orbital inclinations relative to
the line of sight considered (from $0$, $15$, and $18^\circ$) is
constrained by the finite number of test particles used
in our computations and by the necessity of avoiding
great statistical fluctuations in the particle column
density at higher inclinations i. As was said above,
to suppress the fluctuations, the current values of
$n(t)$ for the chosen phase intervals were summed over
$50$ revolutions and were then averaged and smoothed.

We see from Figs. 4–6 that the light curves depend
significantly not only on the orbital inclination to the
line of sight but also on the position of the line of
nodes. The contribution to the extinction from the
CB-disk matter at different distances from its edge
and from the stream propagating from the CB disk
toward the secondary component (it is clearly seen
in Figs. 1 and 2) changes with these parameters. As
a result, light curves differing significantly in both
amplitude and shape are obtained for different binary
orientations. The strong influence of the orientation
of the apsidal line relative to the observer on the light
curves is caused by an azimuthal asymmetry of the
inner CB disk gap. It is the CB-disk matter along
the line of sight that causes a prolonged eclipse of
the central star. The secondary, more compact minimum
is produced by the absorption of radiation in
the stream. The combined action of these two effects
gives rise to two-component light curves (Figs. 4
and 5).

Situations similar to that described above are also
possible in models with coplanar orbits (see Demidova
et al. 2010a, 2010b). However, as our comparative
analysis showed, other things being equal, the brightness
modulation depth in models with noncoplanar
CB disk and orbit is appreciably larger than that in
coplanar binary systems. As a result, a noticeable (in
amplitude) brightness modulation in a young star can
be produced even by the motion of a companion with
a mass corresponding to the mass of a giant planet in
a noncoplanar orbit around it (Fig. 7).
\section{DISCUSSION AND CONCLUSIONS}

The above computations supplement the results
obtained in our previous papers. They show what
types of brightness oscillations may be expected in
young binary systems with different orbital parameters
and different companion masses. In all these
models, the brightness variations are caused by
periodic extinction variations and, therefore, depend
strongly on the orbital inclination and orientation
with respect to the observer. At low inclinations, a
periodic brightness modulation can be observed even
at very low accretion rates, $10^{-10} M_\odot$/yr, typical
of the final evolutionary stages of circumstellar disks.

In this case, a low-mass companion like a brown
dwarf or even a giant planet can be the source of the
perturbations causing the extinction variations.
In our previous papers, we have already pointed
out a great similarity of the model light curves with
the activity cycles observed in UX Ori stars. These
cycles have a wide variety of shapes. Simple cycles
that can be satisfactorily described by a single sine
wave are commonly observed (see, e.g., Artemenko
et al. 2010). In some cases, the cycles are more
complex in shape. For example, the photometric cycle
of BF Ori about 11 yr in duration has a two component
structure (Grinin et al. 1998)\footnote{The reality of this cycle is confirmed by the analysis of the
photometric observations performed with the ASAS telescope
and published in the catalog by Poimanski (2002).} that resembles
the model light curves with two successive
minima (Figs. 4–7). The activity cycle of CO Ori,
which, according to Rostopchina et al. (2007), has
been observed in this star for several decades, also
has a similar shape. From the viewpoint of the models
considered above, the two types of photometric cycles
(simple and two-component) do not differ fundamentally.
They can be obtained in terms of the same model
with different orientations of the orbit relative to the
observer. This peculiarity of the model we considered
is its important advantage over other mechanisms
of cyclic activity in UX Ori stars discussed in the
literature.

\section{ACKNOWLEDGMENTS}

This work was supported by the “Origin and Evolution
of Stars and Galaxies” Program of the Presidium
of the Russian Academy of Sciences and the
Program for Support of Leading Scientific Schools
(NSh-3645.2010.2).
\clearpage

\begin{center}
{\Large\bf REFERENCES}
\end{center}
1. S. A. Artemenko, K. N. Grankin, and P. P. Petrov, Astron. Zh. (2010, in press).\\
2. P. Artymowicz and S. H. Lubow, Astrophys. J. 467, L77 (1994).\\
3. O. Yu. Barsunova, V. P. Grinin, and S. G. Sergeev, Astrofizika 48, 5 (2005).\\
4. M. R. Bate and I. A. Bonnell, Mon. Not. R. Astron. Soc. 285, 33 (1997).\\
5. C. J. Burrows, J. E. Krist, K. R. Stapelfeldt and WFPC2 InvestigationDefinition Team,Bull. Am. Astron.
Soc. 27, 1329 (1995).\\
6. R. E. Cohen, W. Herbst, and E. L. Willams, Astrophys. J. 596, L243 (2003).\\
7. T. V. Demidova, Astrofizika 52, 623 (2009).\\
8. T. V. Demidova, N. Ya. Sotnikova, and V. P. Grinin, Pis’ma Astron. Zh. 36, No. 6, 445 (2010a) [Astron. Lett. 36,
No. 422 (2010)].\\
9. T. V. Demidova, V. P. Grinin, and N. Ya. Sotnikova, Pis’ma Astron. Zh. 36, No. 7, 526 (2010b) [Astron. Lett. 36,
No. 7, 498 (2010)].\\
10. V. P. Grinin, A. N. Rostopchina, and D. N. Shakhovskoi, Pis’ma Astron. Zh. 24, 925
(1998) [Astron. Lett. 24, 802 (1998)].\\
11. V. P. Grinin, E. Stempels, G. Gahm, et al., Astron. Astrophys. 489, 1233 (2008).\\
12. J. D. Larwood and J. C. P. Papaloizou, Mon. Not. R. Astron. Soc. 285, 288 (1997).\\
13. D. Mouillet, J. D. Larwood, J. C. B. Papaloizou, and A. M. Larange, Mon. Not. R. Astron. Soc. 292, 896
(1997).\\
14. A. Natta and B.Whitney, Astron. Astrophys. 364, 633 (2000).\\
15. S. Nordhagen, W. Herbst, E. C. Willams, and
E. Semkov, Astrophys. J. 646, L151 (2006).\\
16. G. Pojmanski, Acta Astron. 52, 397 (2002).\\
17. A. N. Rostopchina, Astron. Zh. 76, 136 (1999) [Astron. Rep. 43, 113 (1999)].\\
18. A. N. Rostopchina, V. P. Grinin, D. N. Shakhovskoi,
et al., Astron. Zh. 84, 60 (2007) [Astron. Rep. 51, 55 (2007)].\\
19. B. A. Smith and R. J. Terrile, Science 226, 1421 (1984).\\
20. N. Ya. Sotnikova, Astrofizika 39, 259 (1996).\\
21. N. Ya. Sotnikova and V. P. Grinin, Pis’ma Astron. Zh.
33, 667 (2007) [Astron. Lett. 33, 594 (2007)].\\

Translated by N. Samus’

\newpage

\clearpage
\begin{figure}[!h]\begin{center}
  \makebox[0.6\textwidth]{\includegraphics[scale=1.05]{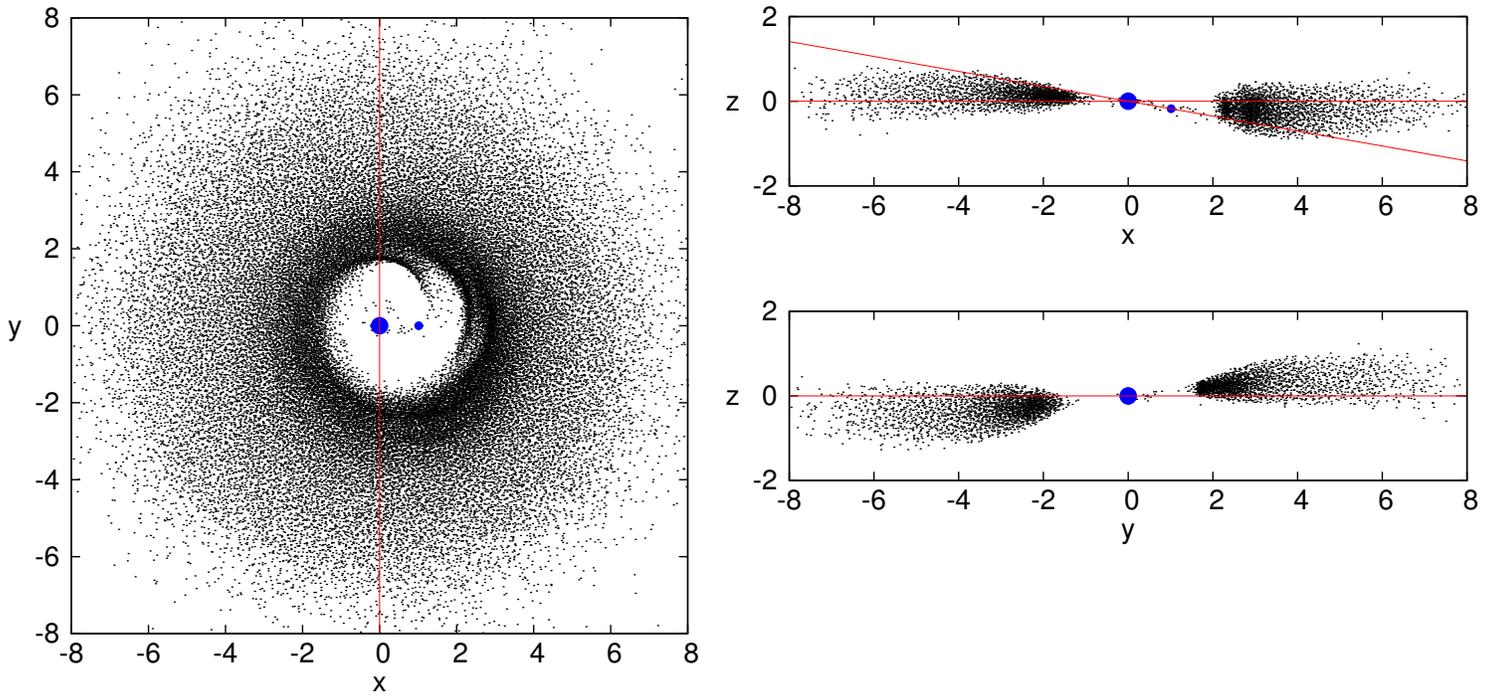}
  }
 \caption{Distribution of matter in model 1 ($q = 0.1$): Left -- top view; right -- disk sections in the XZ and YZ planes. The scale along the
axes is given in units of the orbital semimajor axis. The line of nodes coincides with the Y axis. The binary rotates clockwise}
 \label{disk1}
\end{center}
\end{figure}
\clearpage
\begin{figure}[!h]\begin{center}
  \makebox[0.6\textwidth]{\includegraphics[scale=1.05]{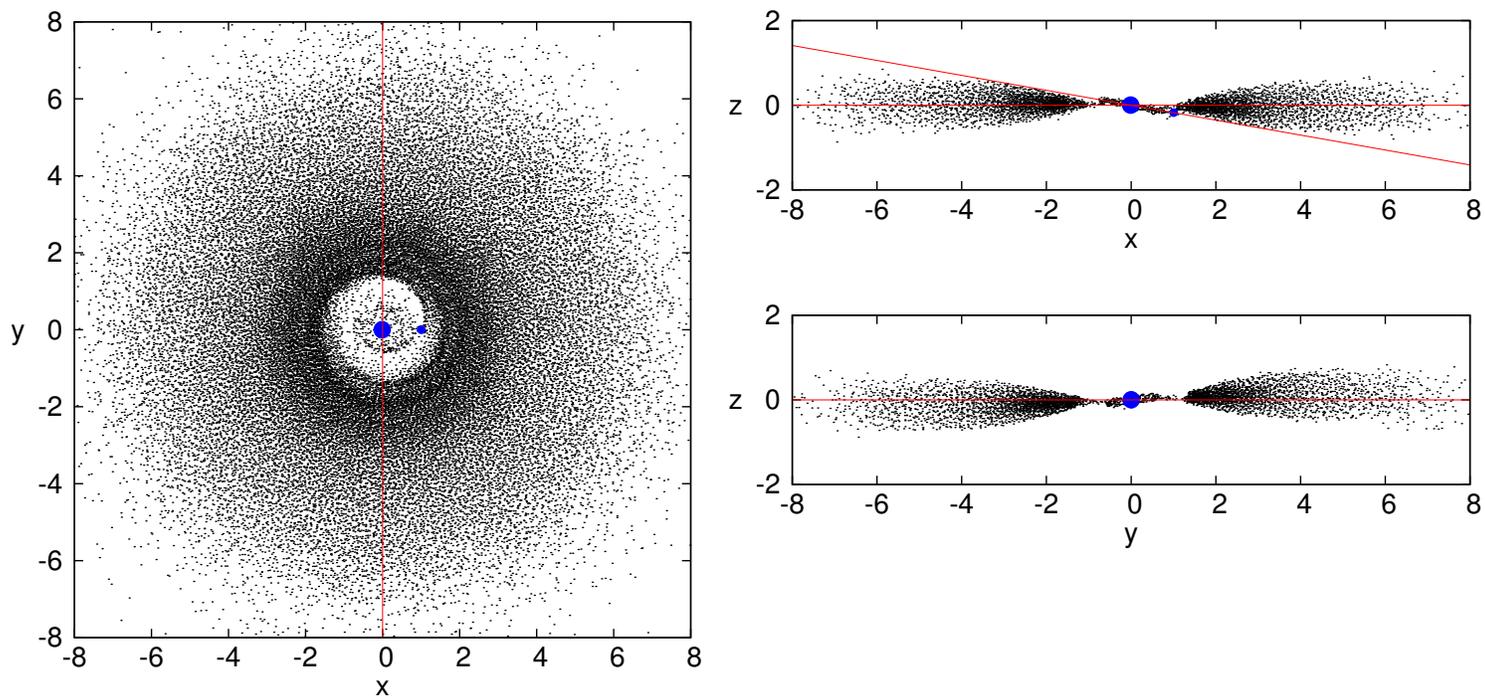}
  }
 \caption{Same as Fig. 1 for model 2  ($q$ = 0.01).}
 \label{disk2}
\end{center}
\end{figure}
\clearpage
\begin{figure}[!h]\begin{center}
\makebox[0.6\textwidth]{\includegraphics[scale=1.0]{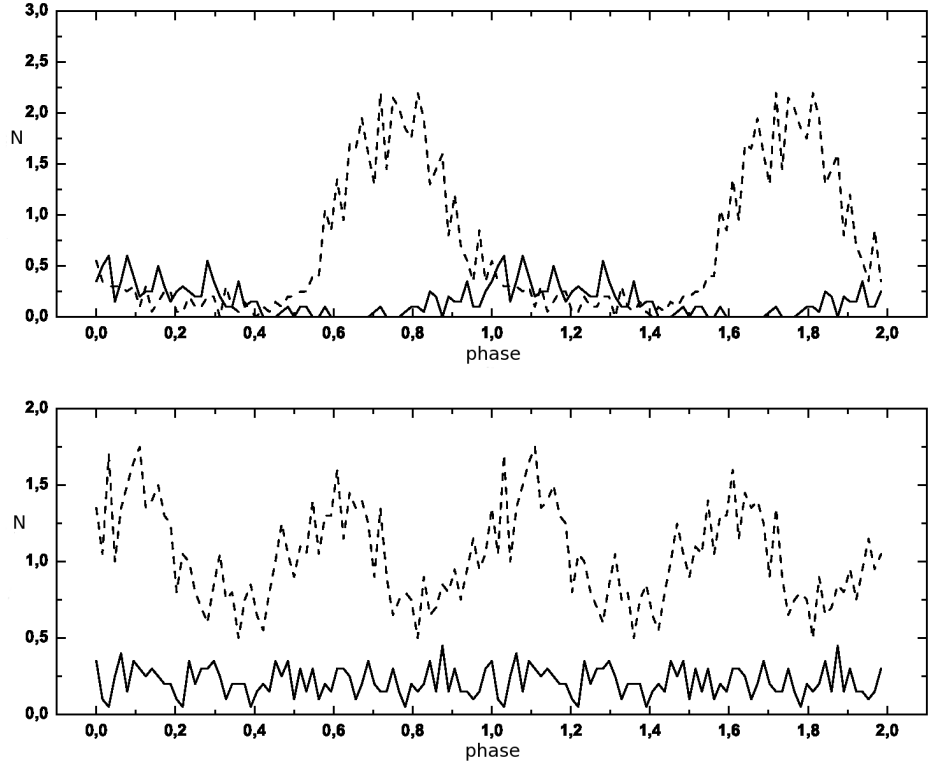}
}
 \caption{ Phase dependences of the accretion rate onto the primary (solid lines) and secondary (dashed lines) components of the
binary: top -- model 1 and bottom -- model 2. The scale along the vertical axis gives the number of test particles accreting onto the binary
components in 1/64 of the orbital period.}
 \label{ar}
\end{center}
\end{figure}
 \clearpage
\begin{figure}\begin{center}
\makebox[0.6\textwidth]{\includegraphics[scale=0.8]{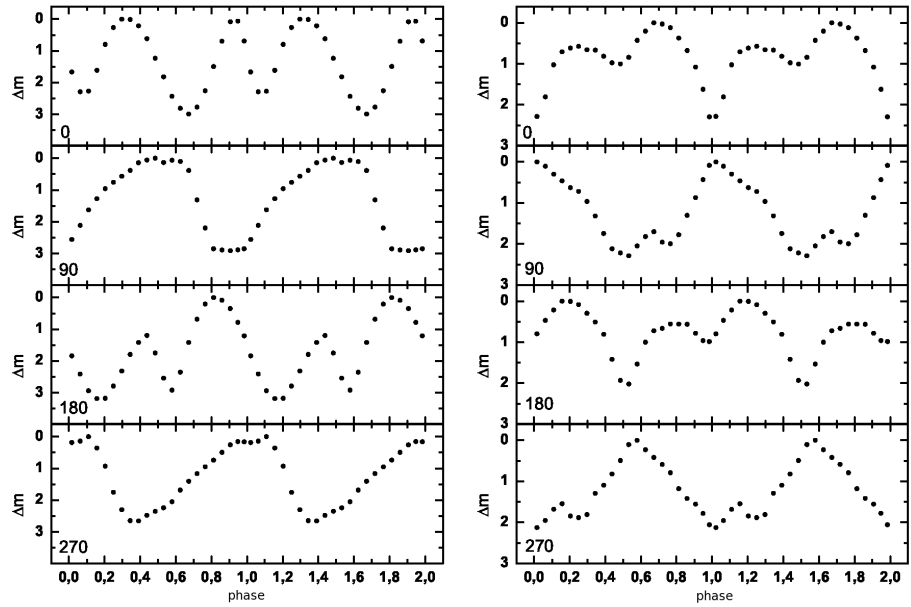} }
 \caption{ Left - light curves in model 1 ($q$ = 0.1) for four orientations of the line of nodes
 ($\phi$ = 0,90,180 and 270$^\circ$) at $i$ =0; right - the same formodel 2($q$ = 0.01). In both cases $\dot M_a = 10^{-9}M_\odot$/yr.}
 \label{lc1ab}
\end{center}
\end{figure}
\clearpage
\begin{figure}[!h]\begin{center}
\makebox[0.6\textwidth]{\includegraphics[scale=0.7]{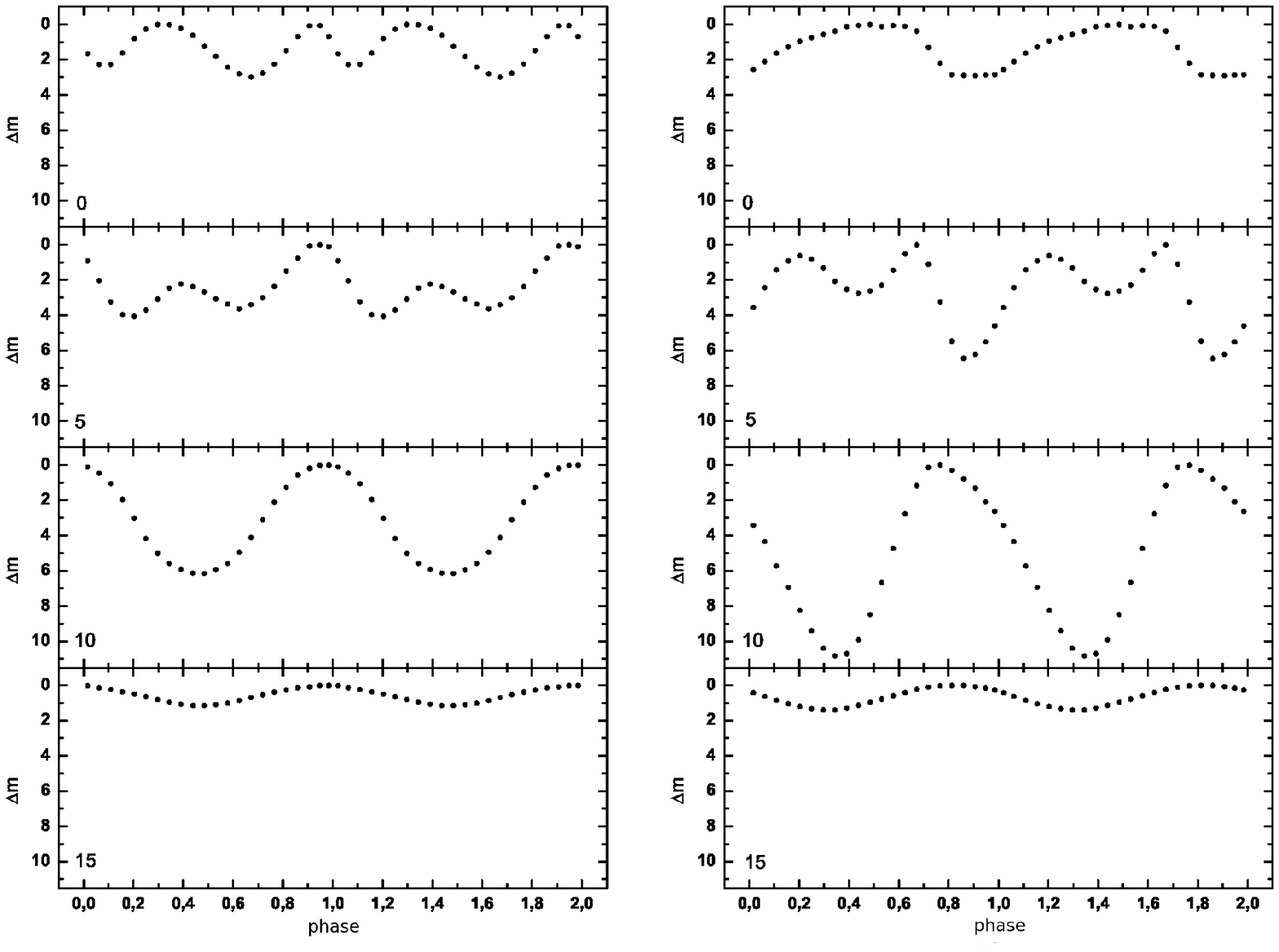}
  }
 \caption{Light curves in model 1 ($q$ = 0.1; $\dot M_a$ = 10$^{-9}M_\odot$/yr)
 for various inclinations of the line of sight to the orbital plane
 and two positions of the line of nodes: left - $\phi$ = 0, right - $\phi$ = 90$^\circ$.}
 \label{lc2a}
\end{center}
\end{figure}
%
\begin{figure}[!h]\begin{center}
  \makebox[0.6\textwidth]{\includegraphics[scale=0.7]{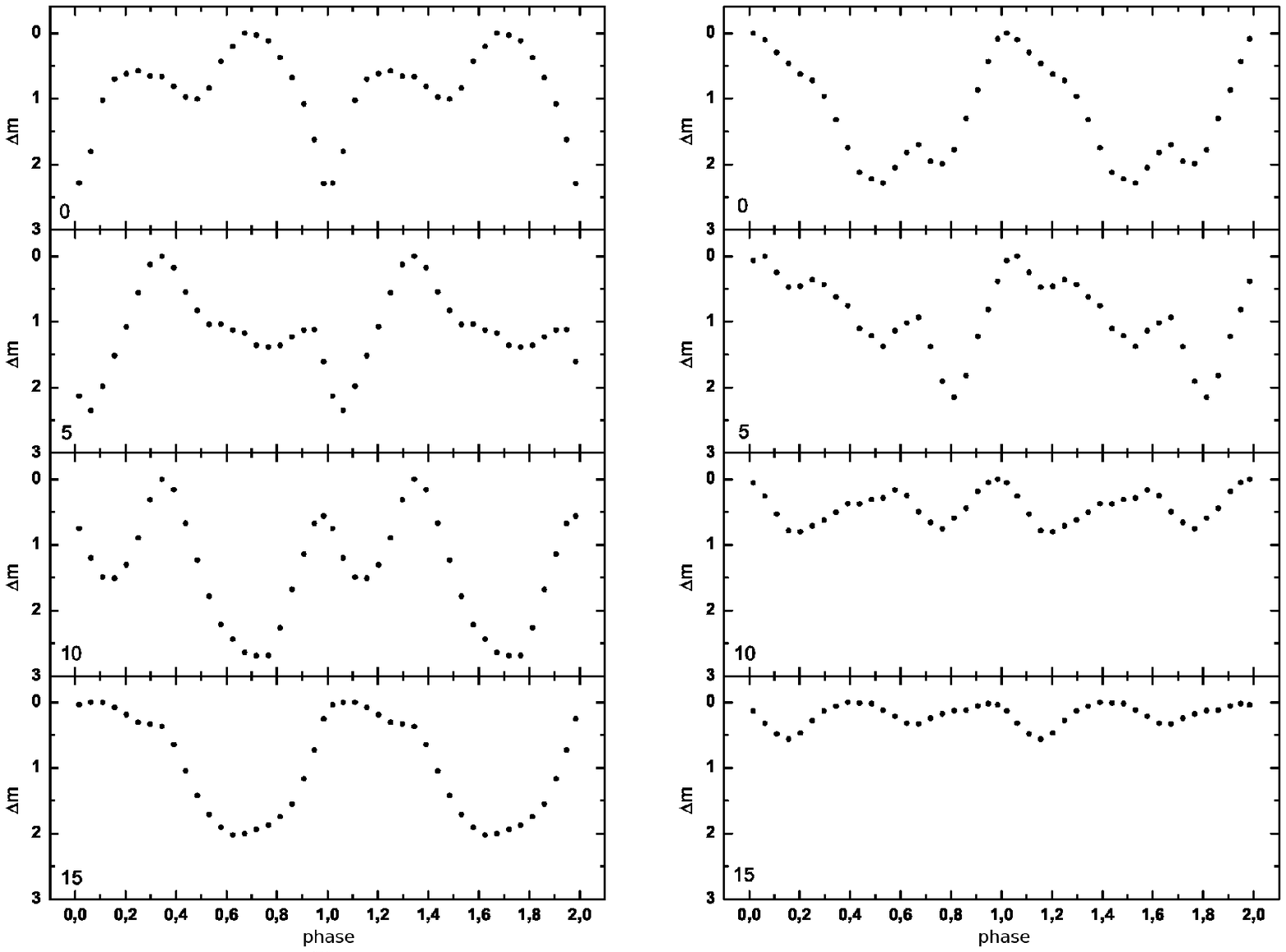}
  }
 \caption{Light curves in model 2 ($q$ = 0.01; $\dot M_a$ = 10$^{-9}M_\odot$/yr)
 for various inclinations of the line of sight to the orbital plane
 and two positions of the line of nodes: left -- $\phi$ = 0, right -- $\phi$ = 90$^\circ$.}
 \label{lc2b}
\end{center}
\end{figure}
\clearpage
\begin{figure}[!h]\begin{center}
\makebox[0.6\textwidth]{\includegraphics[scale=0.7]{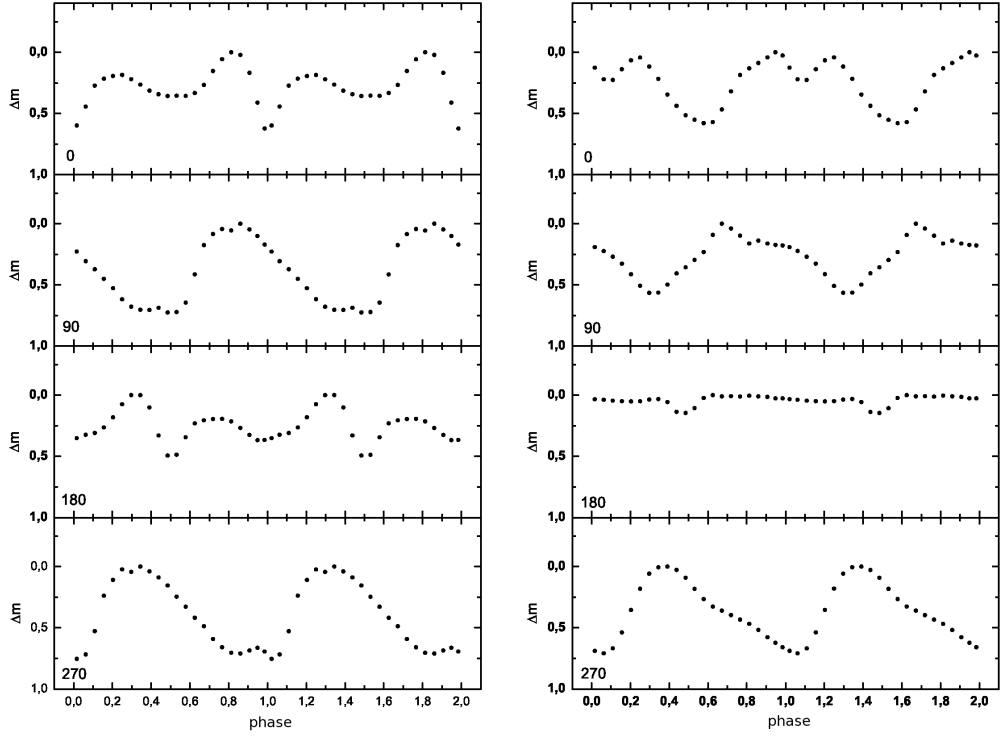} }
\caption{Light curves in model 3 ($q$ = 0.003; $\dot M_a$ =
10$^{-10}M_\odot$/yr) for two inclinations of the orbital plane 
to the line of sight: left -- $i$ = 0, right -- $i$ = 5$^\circ$ 
and four positions of the line of nodes: $\phi$ = $0$, $90$, $180$ \& $270^\circ$.}
 \label{lc3}
\end{center}
\end{figure}

\end{document}